\documentclass[twocolumn,showpacs,preprintnumbers,draftclsnofoot]{revtex4}
\usepackage{amsmath}
\usepackage{amssymb}
\usepackage{graphicx}
\usepackage{dcolumn}
\usepackage{bm}
\usepackage{amssymb}
\usepackage{graphicx}
\usepackage{dcolumn}
\usepackage{bm}

\begin{document}

\title{ Bistability of zigzag edge magnetism in graphene nanoribbons induced by electric field }
\author{Ma Luo\footnote{Corresponding author:swym231@163.com} } 
\affiliation{School of Optoelectronic Engineering, Guangdong Polytechnic Normal University, Guangzhou 510665, China}

\begin{abstract}

In the presence of the Hubbard interaction, graphene zigzag nanoribbons have spontaneous edge magnetism with anti-parallel configuration, whose amplitude can be tuned by a transversal electric field. As the electric field increases or decreases across a critical value, the edges are demagnetized or re-magnetized, respectively. A magnetic field at each edge determines the orientation of the re-magnetization. Thus, a combination of slowly varying transversal electric field and magnetic field in monolayer graphene zigzag nanoribbon could drive the quantum system into a bistability loop. The same phenomenon can be induced in a bilayer/monolayer zigzag nanoribbon without the magnetic field, because the non-symmetry superexchange interaction controls the orientation of the re-magnetization. By this way, the quantum system is switched between ground state and quasi-stable excited state with different magnetism, band structures and conductance. This feature could be used to develop graphene-based spintronic nano-devices without magnetic field.

\end{abstract}

\maketitle

\section{Introduction}

Zigzag nanoribbons of graphene are applicable candidates as integrable spintronic devices \cite{Zutic04,WHan14}, which could reduce the Joule heating. Edge transport of the zigzag nanoribbons could be robust because of the topological properties of the edge states \cite{YuguiYao2011,Motohiko12,YRen16}. In the presence of Hubbard interaction, the zigzag edge host spontaneous magnetism \cite{Mitsutaka96,Hikihara03,Yamashiro03,YoungWooSon06,YoungWoo06,Pisani07,Wunsch08,FernandezRossier08,Jung09,Rhim09,Lakshmi09,Jung09a,Yazyev10,Hancock10,Jung10,Manuel10,Feldner11,DavidLuitz11,JeilJung11,Culchac11,Schmidt12,Soriano12,Karimi12,Schmidt13,Golor13,Bhowmick13,FengHuang13,Ilyasov13,Carvalho14,Lado14,MichaelGolor14,PrasadGoli16,Baldwin16,Ortiz16,Hagymasi16,Ozdemir16,Friedman17,ZhengShi17,XiaoLong18}. The magnetic moment in each zigzag edge is due to uneven population of spin up and down electron at the zigzag edge states. In a narrow zigzag nanoribbon, the edge magnetism at the two zigzag edges interact with each other by superexchange interaction \cite{Jung09a}. In the additional presence of SOCs, the edge magnetism modifies the topological phase diagram, which in turn changes the properties of the topological edge states \cite{maluo2020}. Recently, experimental fabrication of stable zigzag nanoribbons \cite{Ruffieux16} and measurement of the edge magnetization \cite{ZsoltMagda14,MichaelSlota18} make the application of such systems more feasible. By further engineering the nano-structure with graphene, varying type of logical devices have been fabricated, such as graphene-based magneto-logic gate \cite{YanpingLiu20}. The graphene-based spintronic switching devices can be directly connected by carbon-based inter-connecter, such as carbon nanotube \cite{Friedman17}. The network of such devices can largely reduce the energy consumption, and implement large-scale integration.

Logical spintronic devices have been proposed, based on the feature that the conductance of the zigzag nanoribbon is dependent on the configuration of the edge magnetism \cite{Soriano12,Ortiz16,WangYangYang15}. By switching the magnetic moments at the two zigzag edges between being anti-parallel and being parallel, the band structure becomes gapped and gapless, respectively. The critical magnetic field that switches the configuration of the edge magnetism is around 200 T at room temperature \cite{MuonzRojas09}, which become obstacle on the path to applying this system in realistic device. On the other hand, transversal electric field induces imbalance magnetism between two zigzag edges, which drives the zigzag nanoribbon into half metallic phase \cite{YoungWooSon06}. Application of this feature in spin valve has been proposed \cite{MinZhou20}.

We proposed to combine the transversal electric field and a small magnetic field to switch the zigzag nanoribbon between the ground state and the quasi-stable excited state. Iteration solver based on mean field approximation and quasi-static approximation of the tight binding model is applied to studied the evolution of the quantum state as the external electric and magnetic field change with infinitely slow speed. When the transversal electric field slowly increases and exceeds a critical value, the zigzag edges are de-magnetized. After the de-magnetization, slowly decreasing the transversal electric field across the critical value allow the re-magnetization of the two zigzag edges. The direction of the magnetization at each zigzag edge can be controlled by the local magnetic field.

In addition, we proposed a structure of bilayer/monolayer zigzag nanoribbon, in which the role of the small magnetic field is replaced by the superexchange interaction. There are four zigzag edges in the proposed system, so that the number of non-equivalent configuration of edge magnetism is larger than that of monolayer zigzag nanoribbon. As the transversal electric exceed the critical value, the zigzag edge at the bilayer/monolayer interface is not de-magnetized, while those at the open boundaries are de-magnetized. The superexchange interaction between the zigzag edge an the other three zigzag edges at the open boundaries determines the orientation of the re-magnetization. As the transversal electric field slowly varies and alternately across the critical values at positive and negative directions, the quantum system is driven into a bistability loop. Thus, only transversal electric field is required to switch the systems between the ground state and the quasi-stable excited state. The scheme to implement electric control of edge magnetism in carbon based nano-structures without multiferroic materials \cite{Eerenstein06} could bring vast application potential for carbon-based integrated spintronic.

This article is organized as following: In section II, the tight binding model with Hubbard interaction and the simulation methods are reviewed. In section III, the evolution of monolayer graphene zigzag nanoribbon with combination of transversal electric field and magnetic field is studied. In section IV, the static band structure and the evolution of the bilayer/monolayer zigzag nanoribbon with transversal electric field are studied. In section V, the conclusion is given.

\section{Theoretical method}

Assuming that the nanoribbon lays on the x-y plane with the longitudinal axis along the y axis and the width direction along the x axis. The zigzag edges are along the y direction. The tight binding model with Hubbard interaction is given as
\begin{eqnarray}
H=-\sum_{\langle i,j\rangle,\sigma,\kappa}{t_{ij}c_{i,\sigma,\kappa}^{\dag}c_{j,\sigma,\kappa}}-t_{\bot}\sum_{\langle i\kappa,j\bar{\kappa}\rangle,\sigma}{c_{i,\sigma,\kappa}^{\dag}c_{j,\sigma,\bar{\kappa}}} \nonumber \\ +V\sum_{i,\sigma,\kappa}{\kappa c_{i,\sigma,\kappa}^{\dag}c_{i,\sigma,\kappa}}-|e|E_{t}\sum_{i,\sigma,\kappa}{(x_{i}-x_{c}) c_{i,\sigma,\kappa}^{\dag}c_{i,\sigma,\kappa}}\nonumber \\+\mu_{B}B^{z}_{i}\sum_{i,\sigma,\kappa}{\sigma c_{i,\sigma,\kappa}^{\dag}c_{i,\sigma,\kappa}}+U\sum_{i,\kappa}{n_{i,\sigma,\kappa}n_{i,\bar{\sigma},\kappa}}
\end{eqnarray}
where $t_{ij}$ ($t_{\bot}$) is the hopping parameter between the intra-layer (inter-layer) nearest neighbor lattice sites, $2V$ is the inter-layer potential difference due to the vertical gate voltage, $E_{t}$ is the transversal electric field along the width direction ($\hat{x}$ direction), $B^{z}_{i}$ is the out-of-plane direction magnetic field at the $i$-th site, $U$ is the strength of the Hubbard interaction, $i$ and $j$ are the lattice indices of each layer, $\kappa=\pm1$ represents the top and bottom layers, $\sigma=\pm1$ represents spin up and down, $\bar{\kappa}=-\kappa$ and $\bar{\sigma}=-\sigma$. $\mu_{B}=0.5788\times10^{-4}eV\cdot T^{-1}$ is the Bohr magneton, and $\mu_{B}B^{z}_{i}$ is the Zeeman energy splitting. The summation of the first term cover the intra-layer nearest neighbor lattice sites; that of the second term cover the inter-layer nearest neighbor lattice sites. The operator $c_{i,\sigma,\kappa}^{\dag}$ ($c_{i,\sigma,\kappa}$) is the creation (annihilation) operator of the $\pi$ electron on the $i$-th lattice site of the $\kappa$ layer and $\sigma$ spin, and $n_{i,\sigma,\kappa}=c_{i,\sigma,\kappa}^{\dag}c_{i,\sigma,\kappa}$ is the number operator. In the presence of the magnetic field, the hopping parameter is given as $t_{ij}=t_{0}e^{i2\pi\int_{\mathbf{r}_{i}}^{\mathbf{r}_{j}}{\mathbf{A}\cdot d\mathbf{r}}/\Phi_{0}}$, where $\Phi_{0}=\pi\hbar/e$ is the magnetic flux quanta. Two types of magnetic field is considered: (i) for uniform magnetic field $\mathbf{B}^{u}$ with $B^{z}_{i}=B^{z}$, the vector potential is  $\mathbf{A}=(x-x_{mid})B^{z}\hat{y}$; (ii) for linearly varying magnetic field $\mathbf{B}^{l}$ with $(x-x_{mid})B^{z}/W$, the vector potential is $\mathbf{A}=(x-x_{mid})^{2}B^{z}\hat{y}/(2W)$, where $x_{mid}$ is the x-coordinate of the axis in the middle of the nanoribbon, and $2W$ is the width of the nanoribbon. In our calculation, we assume the parameters as $t_{0}=2.8$ eV, $t_{\bot}=0.39$ eV, and $U=t$. For monolayer zigzag nanoribbon, the second and third summation are erased.

The Hubbard interaction induces edge magnetization at the zigzag terminations and quantum fluctuation. The former can be modeled by the mean field theory, while the description of the latter requires more comprehensive method, such as quantum Monte Carlo. For realistic graphene nanoribbons with parameter $U/t\approx1$, comparison between the mean field theory and the quantum Monte Carlo method showed that the effect from the quantum fluctuation can be neglected \cite{Golor13}. By applying the mean field approximation, the Hubbard interaction is approximated as
\begin{equation}
U\sum_{i,\kappa}{n_{i,\sigma,\kappa}n_{i,\bar{\sigma},\kappa}}\approx U\sum_{i,\kappa}{n_{i,\uparrow,\kappa}\langle n_{i,\downarrow,\kappa}\rangle+n_{i,\downarrow,\kappa}\langle n_{i,\uparrow,\kappa}\rangle}
\end{equation}
where $\langle n_{i,\sigma,\kappa}\rangle$ is the expectation of the number operator. For the system with fixed $V$ and $E_{t}$, the tight binding model is self-consistently solved by iteration. In each iteration step, $\langle n_{i,\sigma,\kappa}\rangle$ is obtained by summing the probability density of all quantum states from the previous iteration step, with the occupation factor given by the Fermi-Dirac function with Fermi energy $E_{F}$ and temperature $T$. We assume room temperature in our numerical calculation. For bilayer/monolayer zigzag nanoribbon, the system breaks particle-hole symmetric, so that the intrinsic Fermi energy is not zero. As a result, in each iteration step, an extra iteration is required to determine the Fermi energy by the condition of total charge conservation. In our calculation, we assume that the whole system is half-filled. The magnetic polarization at each lattice site is obtained as $\langle m_{i,\kappa}\rangle\equiv\langle n_{i,+,\kappa}\rangle-\langle n_{i,-,\kappa}\rangle$. If the initial step have different magnetic polarization at the zigzag terminations, the iterative solutions would converge to different magnetic configurations. The solution with the lowest energy is the ground state, and the other solutions with higher energy are the quasi-stable excited states.

The evolution with slowly varying $E_{t}$ is studied by the iterative method. At first, the ground state or the first quasi-stable excited state with $E_{t}=0$ is obtained by the iterative solver. In each of the following evolution step, $E_{t}$ is changed for a small value $\Delta E_{t}$. According to the quasi-static approximation, the relaxation time of the evolution is much smaller than the physical time between adjacent evolution steps, so that the quantum state in each evolution step can be obtained by fully convergent solution. In each evolution step, the iterative solution start from the initial state that is the convergent solution of the previous evolution step. The convergent solution of the iteration is the quantum state of the current evolution step. The quantum state is dependent on the history of the evolution. Assuming that after $N_{E}$ step of the evolution, the transversal electric field starts to change with opposite sign, i.e. change for the among of $-\Delta E_{t}$. If the magnitude of $N_{E}\Delta E_{t}$ is smaller than a critical value, the system evolves back to the initial quantum state as $E_{t}$ reaches zero. By contrast, if the magnitude of $N_{E}\Delta E_{t}$ is larger than the critical value, the system could evolve to a different quantum state, which have different total magnetic moment and band structure. In realistic experiment or device, the transversal electric field should oscillate with a finite frequency. If the frequency is much smaller than the inter-band transition energy of the zigzag nanoribbon, the quasi-static approximation is valid. For the zigzag nanoribbon in ground state, the energy gap is about $0.1$ eV, so that the  frequency of the oscillating transversal electric field is required to be much smaller than $2.4\times10^{13}$ $Hz$.

\section{Monolayer zigzag nanoribbon with magnetic field}

For monolayer zigzag nanoribbons, the spontaneous magnetism at the two zigzag terminations can be either anti-parallel (AF) or parallel (FM) to each other, which is corresponding to the ground state or the quasi-stable excited state, respectively. The switching between the AF and FM states is numerically simulated for a monolayer zigzag nanoribbon with 40 atoms in one unit cell along the width direction.

\begin{figure}[tbp]
\scalebox{0.58}{\includegraphics{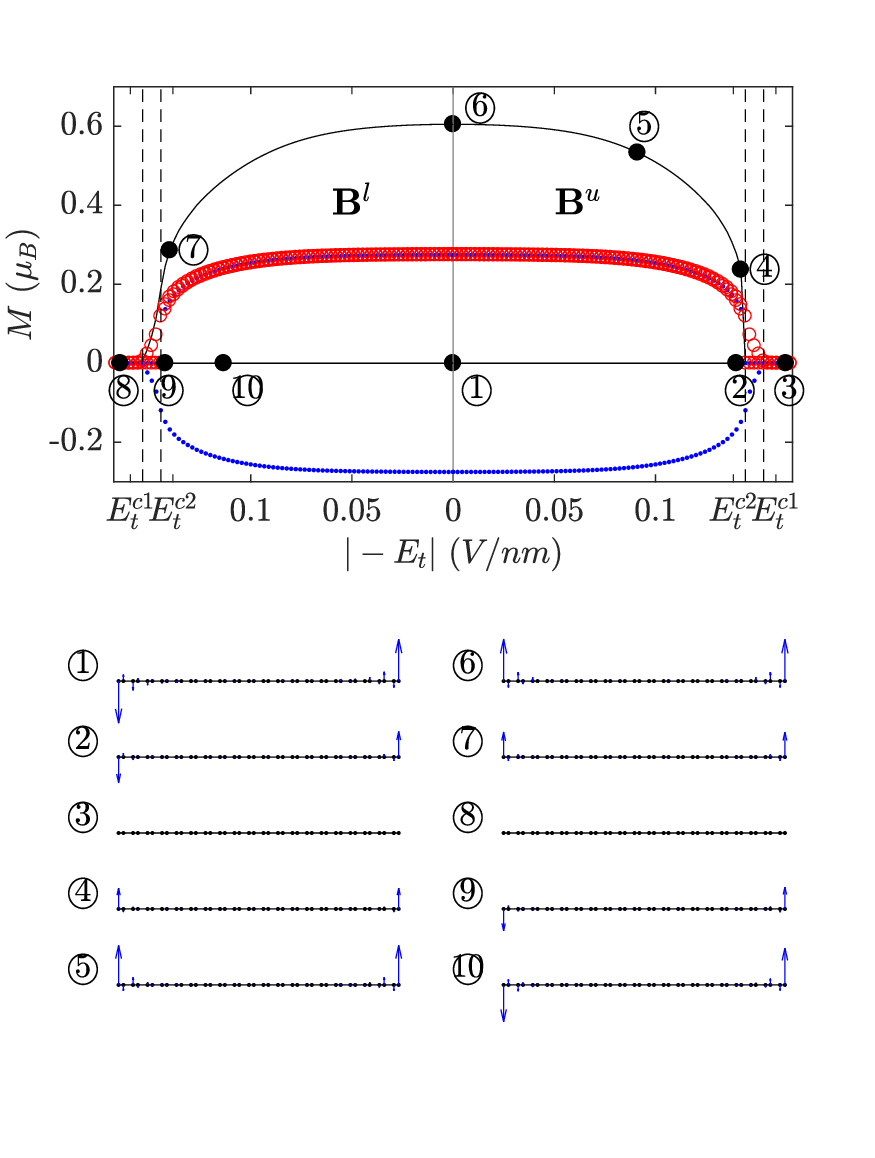}}
\caption{ The bistability evolution of the monolayer zigzag nanoribbon. The total magnetic moment is plotted as black solid line. The magnetic moments at the left and right zigzag terminations are plotted as blue (filled) and red (empty) dots, respectively. The initial state is the AF state. The transversal electric field $E_{t}$ slowly oscillate between $0$ and $0.2$ $V/nm$. The spatially uniform and linearly varying magnetic fields with $B^{z}=10^{-4}$ T are applied for the odd and even periods of the oscillation, respectively. The evolution during the odd and even periods are plotted at the right and left side of the $y$ axis. The quantum state is alternating with the sequence given as $\Large{\textcircled{\small{1}}}\rightarrow\Large{\textcircled{\small{2}}}\rightarrow\cdots\rightarrow\Large{\textcircled{\small{10}}}\rightarrow\Large{\textcircled{\small{1}}}$. The vertical dashed lines marks the critical value of $-E_{t}$. At $E_{t}^{c1}$ and $E_{t}^{c2}$, demagnetization and magnetization of the zigzag terminations occurs, respectively. The distribution of magnetic moment of the ten states in the bistability loop are plotted in the bottom. \label{fig_evolveMonoAF}}
\end{figure}

The bistability loop of the evolution with slowly oscillating transversal electric field is plotted in Fig. \ref{fig_evolveMonoAF}. The magnetic configuration of the initial state is AF. The transversal electric field slowly oscillate between $0$ and $0.2$ $V/nm$. During the first and second periods of the oscillation, spatially uniform and linearly varying magnetic fields with amplitude being $B^{z}=10^{-4}$ T are applied, respectively. The y axis in Fig. \ref{fig_evolveMonoAF} is the total magnetic moment $M$, which is the sum of $\langle m_{i,\kappa}\rangle$ over all lattice sites. The quantum state evolves from $\Large{\textcircled{\small{1}}}$ to $\Large{\textcircled{\small{10}}}$ in sequence, and then circles back to $\Large{\textcircled{\small{1}}}$. The snapshot of the magnetic configurations at the typical steps along the evolution loop, i.e. the quantum states marked as $\Large{\textcircled{\small{1}}}$ to $\Large{\textcircled{\small{10}}}$, are plotted at the bottom part of Fig. \ref{fig_evolveMonoAF}.

For the initial state during the first half of the bistability loop (state $\Large{\textcircled{\small{1}}}$), the population of spin up (down) electron at the right (left) zigzag edge is larger than that of spin down (up) electron, because the corresponding edge band of spin up (down) is below the Fermi level. The magnitude of the magnetic moment at the zigzag terminations is given by the numerical result as $|m^{0}_{Z}|\approx0.275$. As $-E_{t}$ increases ($-E_{t}>0$), charge relaxation occurs due to the tilted local potential, i.e. charge at the right side of the nanoribbon is relaxed to the left side. The spin up electrons at the edge bands of the right zigzag edge are pushed to the left side of the nanoribbon, and filled into the edge bands of the left zigzag edge that is originally above the Fermi level. As a result, the magnetic moment at the two zigzag terminations are slightly decreased, as shown by the state $\Large{\textcircled{\small{2}}}$ in Fig. \ref{fig_evolveMonoAF}. As $-E_{t}$ exceeds a threshold $E_{t}^{c1}$, the local potential at the zigzag terminations overcome the effective exchange fields induced by the spontaneous magnetism, i.e. $W|eE_{t}^{c1}|\approx\frac{f_{c1}}{2}U|\langle m^{0}_{Z}\rangle|$ where $f_{c}$ is a numerical factor that fits the numerical result. At this evolution step, the edge bands of the two spins at the right (left) zigzag edge are both above (below) the Fermi level. Thus, the two zigzag edges are demagnetized, as shown by the state $\Large{\textcircled{\small{3}}}$. In the process of the demagnetization, the unoccupied (occupied) edge bands of the left (right) zigzag edge gradually across the Fermi level. Thus, the magnetic moment at the zigzag terminations gradually reach zero at the threshold, as shown in Fig. \ref{fig_evolveMonoAF}. Before $-E_{t}$ reaches the threshold, $|\langle m^{0}_{Z}\rangle|$ has already been decreased for a small value, so that $f_{c}$ is smaller than one. Numerical result shows that $E_{t}^{c1}=0.1556$ $V/nm$, and then $f_{c1}=0.86$ for this particular case. So far, the effect of the magnetic field is negligible.

After $-E_{t}$ reaching the maximum value, it start to decrease. At this time, the edge bands at the two zigzag edges are nearly two-fold degenerated. Without the edge magnetism, the edge bands are nearly flat. Because of the small external magnetic field, the degeneration is slightly broken. As $-E_{t}$ reaches a threshold $E_{t}^{c2}$, the flat edge bands approach the Fermi level. At this evolution step, the spontaneous magnetization is triggered, so that the magnetic moments at the two zigzag terminations sharply increase, as shown by the state $\Large{\textcircled{\small{4}}}$ in Fig. \ref{fig_evolveMonoAF}. In this system, the local potential at the zigzag terminations and the effective exchange fields are nearly the same, i.e. $W|eE_{t}^{c2}|\approx\frac{f_{c2}}{2}U|\langle m^{0}_{Z}\rangle|$. Numerical result shows that $E_{t}^{c2}=0.1369$ $V/nm$, and then $f_{c2}=0.76$ for this particular case. Because the local external magnetic fields at the two zigzag terminations are the same, the direction of the spontaneous magnetic moments are parallel. As $-E_{t}$ further decrease to zero, the quantum state evolves to the FM state, as shown by the state $\Large{\textcircled{\small{6}}}$ in Fig. \ref{fig_evolveMonoAF}.

The first period of the evolution switches the AF state to the FM state. During the second period of the evolution, the procedure of the evolution is similar to that during the first period of the evolution. As the amplitude of the transversal electric field increases and exceeds the critical value $E_{t}^{c1}$, the two zigzag edges are de-magnetized. After the demagnetization, as the amplitude of the transversal electric field decreases across the critical value $E_{t}^{c2}$, the two zigzag edges are re-magnetized. The direction of the re-magnetization at the two zigzag edges are opposite to each other, because the local magnetic field at the two zigzag edges are opposite. After the re-magnetization, the system evolve to state $\Large{\textcircled{\small{9}}}$ in Fig. \ref{fig_evolveMonoAF}. As the amplitude of the transversal electric field further decreases to zero, the quantum state evolves back to the AF state. The first and second periods of the evolution form the bistability loop.

In summary, the AF and FM states can be switched to each other by applying slowly varying transversal electric field and weak magnetic field, which is feasible in experiment. The conditions to switch the magnetic configurations are summarized in table (\ref{thetable}).  The sign of the transversal electric field is not decisive. As long as the amplitude of the  transversal electric field exceed the critical value, the demagnetization occurs. The directions of the local magnetic field at the two zigzag edges determine the configuration of the re-magnetization, which in turn determine the final state at the time that the transversal electric field decreases to zero.

\section{Bilayer/Monolayer zigzag nanoribbon without magnetic field}

In the previous section, the direction of the magnetization at the zigzag terminations is determined by the local external magnetic field. In the bilayer/monolayer zigzag nanoribbon, the spontaneous magnetism in the middle of the nanoribbon is not demagnetized by the transversal electric field. The superexchange interaction between the zigzag edge and the other three zigzag edges play the role of the magnetic field, so that the external magnetic field is not necessary for switching the quantum state. The process of the switching is described in details as the following.

\subsection{The Static Band Structure}

\begin{figure}[tbp]
\scalebox{0.58}{\includegraphics{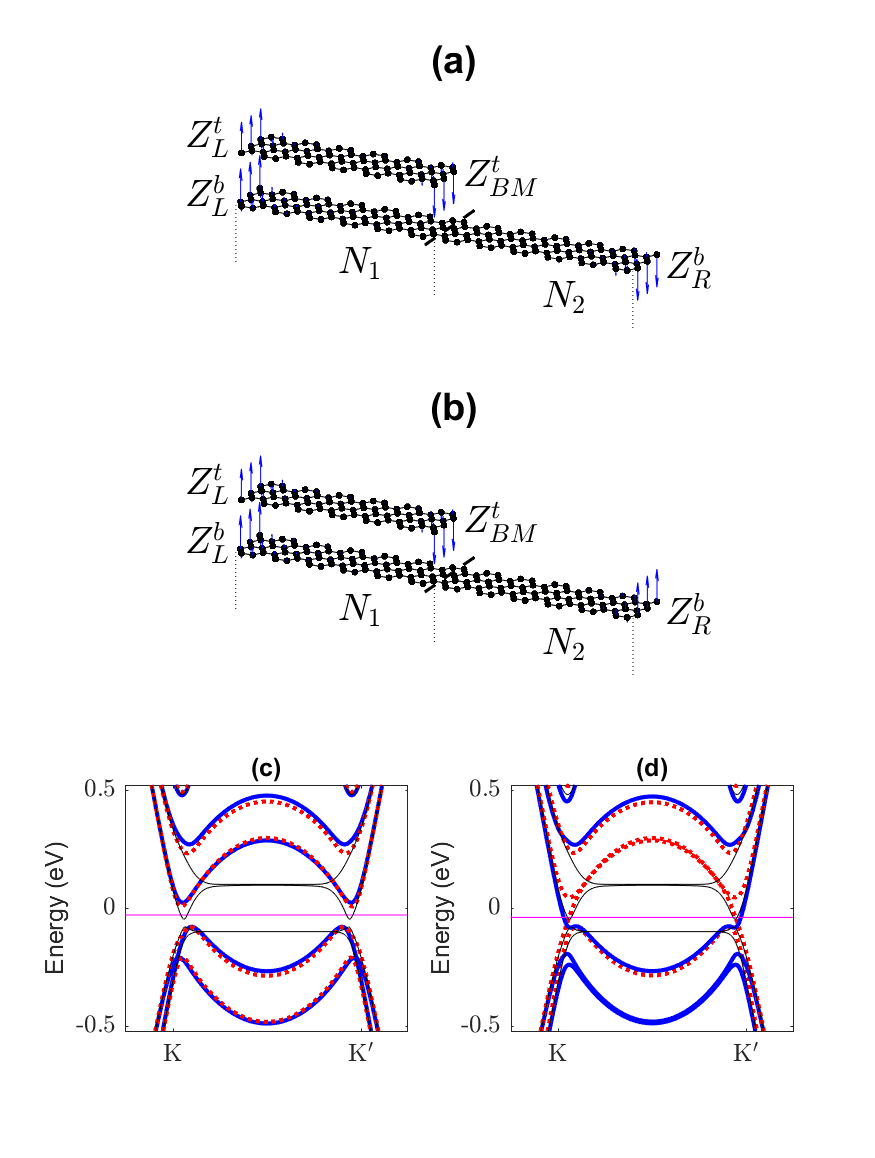}}
\caption{ (a,b) Atomic structure of the bilayer/monolayer zigzag nanoribbon. The numbers of rectangular unit cells along the width direction for the bilayer and the monolayer section are marked on the figures. The structural parameters are $(N_{1}=6, N_{2}=6)$. For the ground state and the first quasi-stable excited state, $\langle m_{i,\kappa}\rangle$ is represented by the arrows in (a) and (b), respectively; the band structures are plotted in (c) and (d), respectively. The bands of spin up and down are plotted as blue (solid) and red (dashed) lines, respectively. The parallel purple (thin) line represents the Fermi level. The system parameters are $V=0.1$ eV and $E_{t}=0$. The bands with $U=0$ is plotted as black (thin) lines are comparison. \label{fig_config1}}
\end{figure}

At first, the ground state and quasi-stable excited states of the system are studied. The structure of the bilayer/monolayer zigzag nanoribbon is plotted in Fig. \ref{fig_config1}. Along the width direction, the bottom layer contains $N_{1}+N_{2}=N$ rectangular unit cells, each of which contains four carbon atoms. The first $N_{1}$ unit cells are covered by the top layer with AB stacking order. The zigzag nanoribbon, designated as $Z_{(N_{1}, N_{2})}$, contains four zigzag edges. We designate the composite index of lattice site $(i,\kappa)$ at each zigzag termination as following: the zigzag terminations at left open boundary of the top and bottom layers as $Z^{t}_{L}$ and $Z^{b}_{L}$, respectively; the zigzag termination at the bilayer/monolayer boundaries as $Z^{t}_{BM}$; the zigzag termination at the right open boundary as $Z^{b}_{R}$.

In the absence of the transversal electric field, all zigzag edges have spontaneous magnetism. The magnetic moment at the termination of each zigzag edge could be either upward or downward. Thus, there are eight nonequivalent magnetic configurations. The band structures of all magnetic configurations with $V=0.1$ eV are calculated by the iterative solver. The magnetic configuration and band structure of the ground state are plotted in Fig. \ref{fig_config1}(a) and (c), respectively. The total energy of the quantum state is decreased, when the magnetic configurations satisfy the following conditions: the magnetic moments at $Z^{t}_{L}$ and $Z^{b}_{L}$ are parallel; the magnetic moments at $Z^{b}_{L}$ and $Z^{b}_{R}$ ($Z^{t}_{L}$ and $Z^{t}_{BM}$) are antiparallel. The ground state satisfy all of the conditions. In the absence of the Hubbard interaction, the edge states form the flat bands at energy $\pm V$, because the states are localized near to the zigzag terminations. The dispersion of the bulk states in the monolayer section is gapless Dirac cone at energy $-V$, but the finite size effect gaps out the band dispersion near to the K and K$^\prime$ points, as shown by the thin black lines in Fig. \ref{fig_config1}(c). In the presence of the Hubbard interaction, the flat bands are bent because of the presence of spatial-dependent effective antiferromagnetic exchange field. The localization of the edge states is weaken by the superexchange interaction, so that the gaps due to finite size effect near to the K and K$^\prime$ points are enlarged, as shown by the thick blue and red lines in Fig. \ref{fig_config1}(c).

The quasi-stable excited states are obtained from the ground state by flipping the magnetic moment at one of the zigzag termination. The interedge superexchange interaction between the magnetic moment at $Z^{b}_{L}$ and $Z^{b}_{R}$ is small because of the large distance between the two edges. Thus, the first quasi-stable excited state is obtained by flipping the magnetic moment at $Z^{b}_{R}$, whose magnetic configuration and band structure are plotted in Fig. \ref{fig_config1}(b) and (d), respectively. The ground state and the first quasi-stable excited state are designated as AF and FM states, because the magnetic moments at the two sides of the bottom  nanoribbon are anti-parallel and parallel, respectively. After flipping the magnetic moment at $Z^{b}_{R}$, a domain wall of the effective antiferromagnetic exchange field is induced in the middle of the nanoribbon. Thus, a pair of chiral edge states for each spin appear, which are gapless at K and K$^\prime$ valleys. For spin up and down, the valley velocities (velocity at K valley minus that at K$^\prime$ valley) are opposite to each other, so that the system hosts dissipationless spin-valley current at the intrinsic Fermi level \cite{KyuWonLee17,Sharma17,Rakhmanov18}. Flipping the magnetic moment at $Z^{b}_{L}$ or $Z^{t}_{L}$ ($Z^{t}_{BM}$) largely increases the energy due to the interedge superexchange interaction between $Z^{b}_{L}$ and $Z^{t}_{L}$ ($Z^{t}_{BM}$ and $Z^{t}_{L}$), so that quasi-stable excited states with much higher energy level are obtained.

\subsection{The Bistability Evolution}

\begin{figure}[tbp]
\scalebox{0.58}{\includegraphics{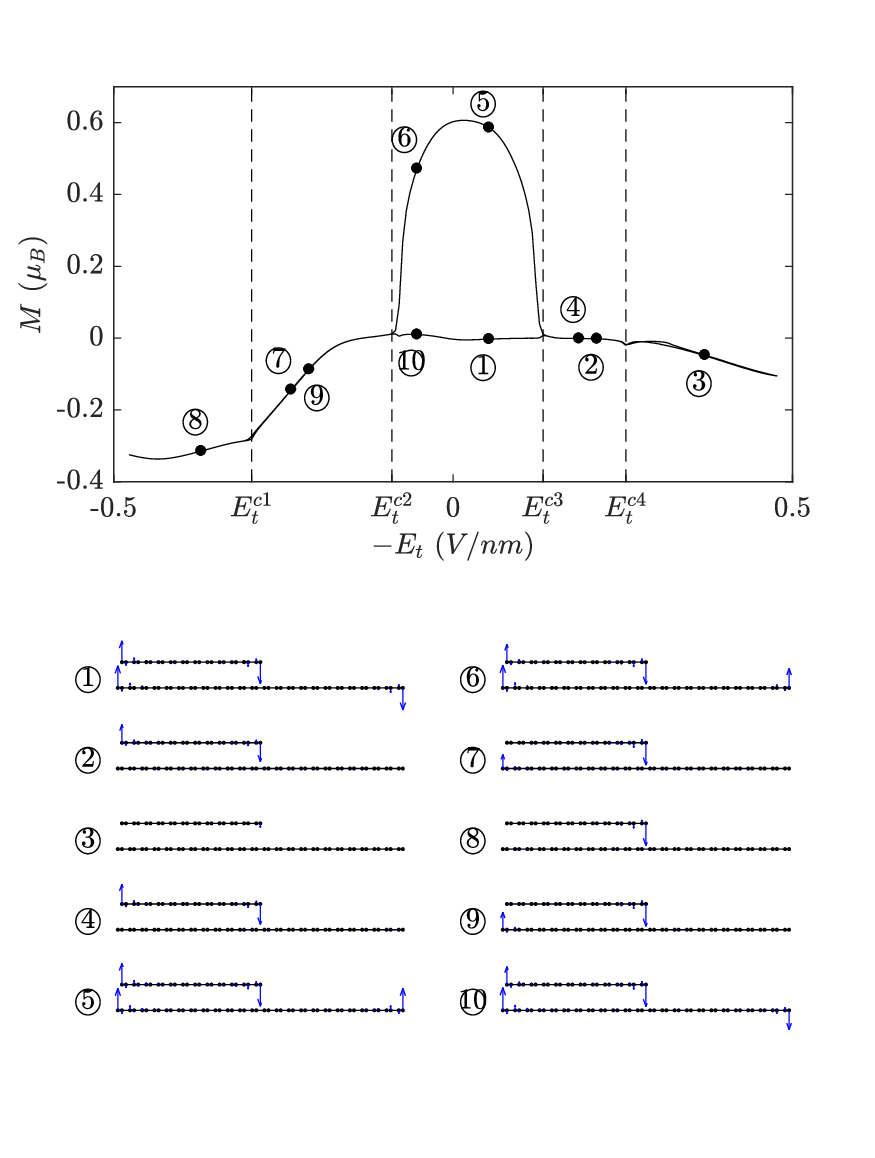}}
\caption{ The bistability evolution of the bilayer/monolayer zigzag nanoribbon. The structural parameters are $(N_{1}=6, N_{2}=6)$. The vertical gate voltage is $V=0.1$ eV. The transversal electric field $-E_{t}$ firstly slowly increase to $+0.48$ $V/nm$, and then slowly oscillates between $\pm0.48$ $V/nm$. The quantum state is alternating with the sequence given as $\Large{\textcircled{\small{1}}}\rightarrow\Large{\textcircled{\small{2}}}\rightarrow\cdots\rightarrow\Large{\textcircled{\small{10}}}\rightarrow\Large{\textcircled{\small{1}}}$. The vertical dashed lines marks the critical value of $-E_{t}$. At $E_{t}^{c1}$, $E_{t}^{c2}$, $E_{t}^{c3}$, $E_{t}^{c4}$, (de)magnetization of the zigzag edge at $Z_{L}^{b}$, $Z_{L}^{t}$ and $Z_{R}^{b}$, $Z_{L}^{b}$ and $Z_{R}^{b}$, $Z_{L}^{t}$ occurs, respectively. The distribution of magnetic moment of the ten states in the bistability loop are plotted in the bottom. \label{fig_evole1}}
\end{figure}

As the transversal electric field firstly slowly increasing to $+0.48$ $V/nm$, and then slowly oscillating between $\pm E_{t0}=0.48$ $V/nm$, the evolution starting from the ground state with $V=0.1$ $eV$ is represented by the evolution loop in Fig. \ref{fig_evole1}. As the system evolves, the quantum state evolves from $\Large{\textcircled{\small{1}}}$ to $\Large{\textcircled{\small{10}}}$ in sequence, and then circles back to $\Large{\textcircled{\small{1}}}$. The snapshot of the magnetic configurations at the typical steps along the evolution loop, i.e. the quantum states marked as $\Large{\textcircled{\small{1}}}$ to $\Large{\textcircled{\small{10}}}$, are plotted at the bottom part of Fig. \ref{fig_evole1}. As $-E_{t}$ acrosses the critical values $E_{t}^{c(1-4)}$, demagnetization (re-magnetization) of certain zigzag edges occurs. The demagnetization (re-magnetization) and the critical value are analyzed as the following.

At the ground state (initial state), the magnetic moments at $Z^{b}_{L}$ and $Z^{t}_{L}$ are antiparallel to those at $Z^{b}_{R}$ and $Z^{t}_{BM}$, so that $M$ is nearly zero. The magnitudes of $\langle m_{i,\kappa}\rangle$ at the zigzag terminations are given by the numerical result as $|\langle m_{Z}^{0}\rangle|\approx0.28$, with $Z\in\{Z^{b}_{L}, Z^{b}_{R}, Z^{t}_{L}, Z^{t}_{BM}\}$ being the index of the zigzag terminations. At $Z^{b}_{L}$ or $Z^{t}_{L}$ ($Z^{b}_{R}$ or $Z^{t}_{BM}$) the populations of spin up (down) electron is larger than that of spin down (up) electron, because the edge band of spin up (down) is below the Fermi level. As $-E_{t}$ increases ($-E_{t}>0$), charge relaxation occurs due to the tilted local potential, i.e. charge at the right side of the nanoribbon is relaxed to the left side. Because of the magnetization at $Z^{b}_{R}$, the spin down electrons at $Z^{b}_{R}$ are pushed to the left side of the nanoribbon. The local potential at $Z^{b}_{L}$ is smaller than that at $Z^{t}_{L}$ due to the vertical gate voltage, so that the spin down electrons are filled into $Z^{b}_{L}$. As a result, $|\langle m^{0}_{Z^{b}_{L(R)}}\rangle|$ are slightly decreased. As $-E_{t}$ exceeds a threshold, the local potential at $Z^{b}_{L}$ and $Z^{b}_{R}$ overcome the effective exchange fields induced by the spontaneous magnetism, i.e. the edge bands of both spin are above and below the Fermi level, respectively. Thus, the two zigzag edges are demagnetized. The threshold is given as
\begin{equation}
W|e|E_{t}^{c3}\approx\frac{f_{c}}{2}U|\langle m^{0}_{Z}\rangle| \label{criticalEc1}
\end{equation}
where $2W=(3N-1)a_{c}$ is the width of the bottom nanoribbon with $a_{c}$ being the bond length, $f_{c}$ is a numerical factor that fits the numerical result. Before $-E_{t}$ reaches the threshold, $|\langle m^{0}_{Z^{b}_{L(R)}}\rangle|$ has already been decreased for a small value, so that $f_{c}$ is smaller than one. The demagnetization can be visualized from the change between the spatial distribution of the magnetic moment at state $\Large{\textcircled{\small{1}}}$ and state $\Large{\textcircled{\small{2}}}$ in Fig. \ref{fig_evole1}. Similarly, demagnetization at $Z^{t}_{L}$ and $Z^{t}_{BM}$ occurs at the critical transversal electric field, which is given as
\begin{equation}
W_{1}|e|E_{t}^{c4}\approx\frac{f_{c}}{2}U|\langle m^{0}_{Z}\rangle| \label{criticalEc2}
\end{equation}
where $2W_{1}=(3N_{1}-1)a_{c}$ is the width of the top nanoribbon. However, $Z^{t}_{BM}$ is not completely demagnetized. As $-E_{t}$ further increase, $|\langle m^{0}_{Z}\rangle|$ at $Z^{t}_{BM}$ slowly increase, as shown in Fig. \ref{fig_evole1}.

In the next stage of the  evolution, $-E_{t}$ slowly decreases (while remaining $-E_{t}>0$). As $-E_{t}$ passes $E_{t}^{c4}$, $Z^{t}_{L}$ and $Z^{t}_{BM}$ are magnetized to the original configuration, because previously $Z^{t}_{BM}$ was not completely demagnetized and the interedge interaction between the two edges favors the antiparallel configuration. As $-E_{t}$ further decreases and passes $E_{t}^{c3}$, $Z^{b}_{L}$ and $Z^{b}_{R}$ are magnetized. $Z^{b}_{L}$ is magnetized to the original direction, because the interedge interaction between $Z^{b}_{L}$ and $Z^{t}_{L}$ favors parallel configuration. The magnetization of $Z^{b}_{R}$ is determined by the competition among three pairs of interedge interactions: $(Z^{b}_{R}\Leftrightarrow Z^{t}_{L})$, $(Z^{b}_{R}\Leftrightarrow Z^{b}_{L})$, and $(Z^{b}_{R}\Leftrightarrow Z^{t}_{BM})$, all of which favor the antiparallel configuration. The interedge interaction $(Z^{b}_{R}\Leftrightarrow Z^{t}_{L})$ is inter-layer with large distance, so that it is the weakest. The interedge interaction $(Z^{b}_{R}\Leftrightarrow Z^{b}_{L})$ is intra-layer with large distance, and the interedge interaction $(Z^{b}_{R}\Leftrightarrow Z^{t}_{BM})$ is inter-layer with small distance. Thus, the strength of the two interedge interactions are similar. In this stage of the evolution, we have $-|e|E_{t}>0$ and $V>0$. By increasing the vertical gate voltage $V$, the difference of local potential between $Z^{b}_{R}$ and $Z^{t}_{BM}$, which is $-|e|E_{t}N_{2}a_{c}-2V$, is decreased. Thus, the interedge interaction $(Z^{b}_{R}\Leftrightarrow Z^{t}_{BM})$ is enhanced. Meanwhile, the vertical gate voltage does not change the interedge interaction $(Z^{b}_{R}\Leftrightarrow Z^{b}_{L})$, because the two edges are at the same layer.  With $V=0.1$ $eV$, the interedge interaction $(Z^{b}_{R}\Leftrightarrow Z^{t}_{BM})$ is larger than the interedge interaction $(Z^{b}_{R}\Leftrightarrow Z^{b}_{L})$. Thus, the direction of the magnetization at $Z^{b}_{R}$ is determined by the interedge interaction $(Z^{b}_{R}\Leftrightarrow Z^{t}_{BM})$. As a result, the magnetic configuration is evolved to state $\Large{\textcircled{\small{5}}}$, instead of returning to state $\Large{\textcircled{\small{1}}}$. Continuing from the quantum state $\Large{\textcircled{\small{5}}}$, as $-E_{t}$ decrease to zero, the system evolves to the first quasi-stable excited state, which have large total magnetic moment. On the other hand, if the vertical gate voltage $V$ is not large enough, the direction of the magnetization at $Z^{b}_{R}$ is determined by the interedge interaction $(Z^{b}_{R}\Leftrightarrow Z^{b}_{L})$. Thus, the magnetic configuration is evolved back to state $\Large{\textcircled{\small{1}}}$.

In the following stage of the evolution, $-E_{t}$ becomes negative with increasing magnitude. Due to the charge relaxation, $|\langle m^{0}_{Z^{b}_{L}}\rangle|$, $|\langle m^{0}_{Z^{t}_{L}}\rangle|$ and $|\langle m^{0}_{Z^{b}_{R}}\rangle|$ are slightly decreased. Because the vertical gate voltage induces positive (negative) local potential at top (bottom) layer, the magnitude of total local potential at $Z^{t}_{L}$ and $Z^{b}_{R}$ are larger than that at $Z^{b}_{L}$. As a result, when $-E_{t}$ reaches the critical value $E_{t}^{c2}$, $Z^{t}_{L}$ and $Z^{b}_{R}$ are demagnetized, while $Z^{b}_{L}$ remain magnetized. The critical value is given as
\begin{equation}
W|e|E_{t}^{c2}\approx V-\frac{f_{c}}{2}U|\langle m^{0}_{Z}\rangle| \label{criticalEc3}
\end{equation}
As the magnitude of $-E_{t}$ further increases, charge relaxation occurs between $Z^{b}_{L}$ and the monolayer section of the nanoribbon. Combining the effect of $V$ and $-E_{t}$, the critical value that $Z^{b}_{L}$ is demagnetized is given as
\begin{equation}
W_{1}|e|E_{t}^{c1}\approx-V-\frac{f_{c}}{2}U|\langle m^{0}_{Z}\rangle| \label{criticalEc4}
\end{equation}
Because $-E_{t}$ does not change the local potential at $Z^{t}_{BM}$, $|\langle m^{0}_{Z^{t}_{BM}}\rangle|$ is hardly changed, as shown by state $\Large{\textcircled{\small{8}}}$ in Fig. \ref{fig_evole1}.

In the last quarter of the evolution, the magnitude of $-E_{t}$ slowly decreases. As $-E_{t}$ reaches $E_{t}^{c1}$, $Z^{b}_{L}$ is magnetized to  the original direction, because the interedge interaction between $Z^{b}_{L}$ and $Z^{t}_{BM}$ favors the antiparallel configuration. As $-E_{t}$ reaches $E_{t}^{c2}$, $Z^{t}_{L}$ and $Z^{b}_{R}$ are magnetized. $Z^{t}_{L}$ is magnetized to the original direction, because the interedge interaction $(Z^{t}_{L}\Leftrightarrow Z^{b}_{L})$ favors the parallel configuration, and the interedge interaction $(Z^{t}_{L}\Leftrightarrow Z^{t}_{MB})$ favors the antiparallel configuration. The magnetization of $Z^{b}_{R}$ is again determined by the competition between the two pairs of interedge interactions: $(Z^{b}_{R}\Leftrightarrow Z^{b}_{L})$, and $(Z^{b}_{R}\Leftrightarrow Z^{t}_{BM})$. In this stage of the evolution, we have $-|e|E_{t}<0$ and $V>0$, so that the vertical gate voltage effectively decreases the interedge interaction $(Z^{b}_{R}\Leftrightarrow Z^{t}_{BM})$. The interedge interaction $(Z^{b}_{R}\Leftrightarrow Z^{b}_{L})$ dominates, so that $Z^{b}_{R}$ is magnetized to have antiparallel configuration with $Z^{b}_{L}$. Thus, the magnetic figuration is evolved to the quantum state $\Large{\textcircled{\small{10}}}$, instead of returning to the quantum state $\Large{\textcircled{\small{6}}}$. As the magnitude of $-E_{t}$ decrease to zero, the system evolves to the ground state. So far, the evolution completes one bistability loop.

\begin{figure}[tbp]
\scalebox{0.58}{\includegraphics{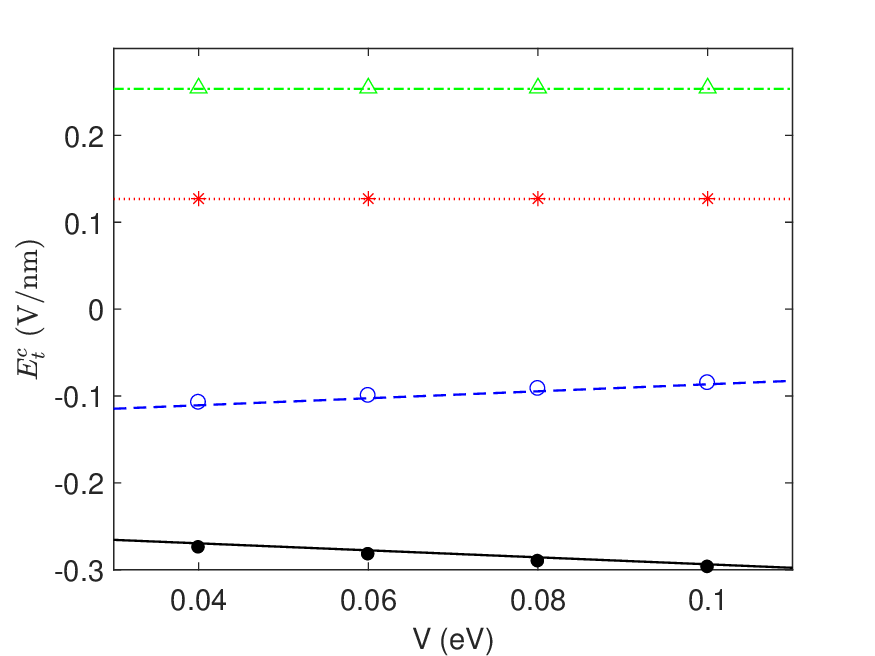}}
\caption{ The critical value of $-E_{t}$ in the bistability loop versus the vertical gate voltage $V$. The structural parameters are $(N_{1}=6, N_{2}=6)$.  Numerical results of $E_{t}^{c1}$, $E_{t}^{c2}$, $E_{t}^{c3}$, $E_{t}^{c4}$ are plotted as black dots, blue empty dots, red stars, green triangles, respectively. The analytical formulas that are fit to the numerical results are plotted as  black (solid), blue (dashed), red (dotted), green (dash-dotted) lines, respectively. \label{fig_evole1C}}
\end{figure}

\subsection{Critical step of the bistability loop}

The two critical steps in the bistability loop are the evolution from state $\Large{\textcircled{\small{4}}}$ to $\Large{\textcircled{\small{5}}}$, and from state $\Large{\textcircled{\small{9}}}$ to $\Large{\textcircled{\small{10}}}$, which lead to the switching of the magnetic configurations AF$\rightarrow$FM and FM$\rightarrow$AF, respectively. The conditions to switch the magnetic configurations are summarized in table (\ref{thetable}). If the maximum magnitude of $-E_{t}$ is smaller than $|E_{t}^{c2}|$ and $|E_{t}^{c3}|$, the evolution always return to the AF state. If the maximum magnitude of $-E_{t}$ is smaller than $|E_{t}^{c1}|$ and $|E_{t}^{c4}|$, but larger than $|E_{t}^{c2}|$ and $|E_{t}^{c3}|$, the  evolution can still enters the bistability loop. According to the description of the magnetization at $Z^{b}_{R}$ in these two critical step, the decisive reason of entering the bistability loop is that the combination of the transversal electric field and the sizable vertical gate voltage changes the competition between the two interedge interactions: $(Z^{b}_{R}\Leftrightarrow Z^{b}_{L})$ and $(Z^{b}_{R}\Leftrightarrow Z^{t}_{BM})$. Because the atomic configuration of the bilayer/monolayer zigzag nanoribbon is not symmetric about the axis, the competition between the two interedge interactions depends on the sign of the transversal electric field. As the transversal electric field decreases amplitude with different sign, the directions of the re-magnetization are different, which lead the evolution to different magnetic configuration. As a result, the periodic evolution enter the bistability loop.

Evolutions with varying $V$ are numerically calculated, which found that $|V|>0.035$ eV is required for entering the bistability loop. The critical value of $-E_{t}$ where the demagnetization occurs versus the vertical gate voltage is extracted from the numerical result, as shown in Fig. \ref{fig_evole1C}. By fitting the analytical formula in Eq. (\ref{criticalEc1}-\ref{criticalEc4}), the numerical factor $f_{c}=0.804$ is obtained. Because of the selective magnetization at the two critical steps, the evolution proceeds along the anticlockwise direction of the bistability loop in Fig. \ref{fig_evole1}. If $-E_{t0}$ firstly decreases to $-0.48$ $V/nm$ and then slowly oscillates between $\pm0.48$ $V/nm$, the first two quarters of the evolution follows the path: $\Large{\textcircled{\small{1}}}\Rightarrow\Large{\textcircled{\small{10}}}\Rightarrow\Large{\textcircled{\small{9}}}\Rightarrow\Large{\textcircled{\small{8}}}\Rightarrow\Large{\textcircled{\small{9}}}\Rightarrow\Large{\textcircled{\small{10}}}\Rightarrow\Large{\textcircled{\small{1}}}$, and returns to the ground state. As $E_{t0}$ continue to oscillate, the following evolution enters the bistability loop along the anticlockwise direction. By contrast, if the vertical gate voltage is reversed, i.e. $V<0.035$ eV is applied, the evolution proceeds along the clockwise direction of the bistability loop in Fig. \ref{fig_evole1}. The ground state and the first quasi-stable excited state are gapped and gapless, respectively, so that the conductance of the nanoribbon is alternatively switched off and on in the bistability loop.


\begin{table}
\centering
\caption{The condition of the critical step in the bistability loop for monolayer and bilayer/monolayer nanoribbon, including the sign of the transversal electric field, minimum amplitude of the transversal electric field, the external magnetic field and the minimum gated voltage. The first three rows and the last three rows are for the monolayer nanoribbon (M-N) and bilayer/monolayer nanoribbon (B/M-N), respectively. \label{thetable}}
\begin{center}
\begin{tabular}{|l|c|c|c|c|}
\hline
M-N  & $sign(-E_{t})$ & $|e|E_{t}^{c}$ & $\mathbf{B}$ & $min(|V|)$ \\ \hline
AF$\rightarrow$FM & $\pm$ & $f_{c1}U|\langle m^{0}_{Z}\rangle|/(2W)$ & $\mathbf{B}^{u}$  & 0  \\ \hline
FM$\rightarrow$AF & $\pm$ & $f_{c1}U|\langle m^{0}_{Z}\rangle|/(2W)$ & $\mathbf{B}^{l}$ & 0 \\ \hline \hline
B/M-N  & $sign(-E_{t})$ & $|e|E_{t}^{c}$ & $\mathbf{B}$ & $min(|V|)$ \\ \hline
AF$\rightarrow$FM & $+$ & $f_{c}U|\langle m^{0}_{Z}\rangle|/(2W)$ & 0 & 0.035 eV \\\hline
FM$\rightarrow$AF & $-$ & $V-f_{c}U|\langle m^{0}_{Z}\rangle|/(2W)$ & 0 & 0.035 eV  \\
\hline
\end{tabular}
\end{center}
\end{table}

\section{Conclusion}

In conclusion, the magnetic configurations of the graphene zigzag nanoribbons can be switched from one to another by slowly varying transversal electric field, which demagnetizes and then re-magnetizes the zigzag edges. For monolayer zigzag nanoribbons, the weak magnetic field determines the direction of the re-magnetization of each zigzag edge, which in turn control the magnetic configuration after the switching. For the bilayer/monolayer zigzag nanoribbons, the magnetic configurations can be switched by solely applying electric field. Because of the asymmetric structure, the combination of the sizable vertical gate voltage and the transversal electric field with different sign induce different inter-edge superexchange interaction. Thus, the sign of the transversal electric field controls the configuration of the re-magnetization. As the transversal electric field slowly oscillates between positive and negative value with sufficient amplitude, the evolution of the quantum system enters a bistability loop, which alternates between ground state and the quasi-stable excited state with different band structure. The feature can be applied in electrically controlled graphene-based spintronic nano-device.

\begin{acknowledgments}
This project is supported by the National Natural Science Foundation of China (Grant:
11704419).
\end{acknowledgments}

\clearpage

\end{document}